\documentclass{iopart}

\usepackage{iopams,setstack}
\usepackage{epsf}
\usepackage{cite}

\newcommand{\bd}{\begin{displaymath}}
\newcommand{\ed}{\end{displaymath}}
\newcommand{\be}{\begin{equation}}
\newcommand{\ee}{\end{equation}}
\newcommand{\ba}{\begin{eqnarray}}
\newcommand{\ea}{\end{eqnarray}}

\begin{document}

\paper{Setting-up tunneling conditions by means of Bohmian mechanics}

\author{A S Sanz and S Miret-Art\'es}

\address{Instituto de F\'{\i}sica Fundamental - CSIC,
Serrano 123, 28006 Madrid, Spain}

\eads{\mailto{asanz@iff.csic.es} and \mailto{s.miret@iff.csic.es}}

\begin{abstract}
Usually tunneling is established after imposing some matching
conditions on the (time-independent) wave function and its first
derivative at the boundaries of a barrier.
Here an alternative scheme is proposed to determine tunneling and
estimate transmission probabilities in time-dependent problems, which
takes advantage of the trajectory picture provided by Bohmian
mechanics.
From this theory a general functional expression for the transmission
probability in terms of the system initial state can be reached.
This expression is used here to analyze tunneling properties and
estimate transmissions in the case of initial Gaussian wave packets
colliding with ramp-like barriers.
\end{abstract}






\section{Introduction}
 \label{sec1}

Quantum tunneling, the possibility for a system to pass from one state
$A$ to another state $B$ {\it through} an energetic barrier, can be
considered as one of the characterizing phenomena of quantum mechanics
---although it is more general, appearing whenever a system is described
by a wave equation and there is coupling among {\it evanescent
waves} \cite{main}, as it happens in optics \cite{born-wolf}, for
example.
In 1928, shortly after the formulation of Schr\"odinger's equation,
tunneling was proposed as the physical mechanism that explained both
field electron emission \cite{fowler} and alpha decay
\cite{gamow,gurney}.
For about 30 years, no satisfactory explanation was possible for these
effects, well-known since the end of the XIXth century
\cite{wood,elster}.
Nowadays tunneling not only
appears in tunnel microscopy or nuclear physics, but also in other
fields, such as semiconductors or catalytic and enzymatic
reactions, with important direct applications. There is so much
written about tunneling that it seems there is few space left for
new conceptual ideas about this phenomenon or mechanism.
Nonetheless, it continues stirring our interest.

In the study and analysis of tunneling problems, many different
techniques (classical, semiclassical and quantum-mechanical) have
been considered in the literature \cite{razavy,ankerhold}. For
example, for low tunneling (i.e., near the top of the barrier),
classical and semiclassical approaches, such as the WKB
approximation, have been shown to be very appropriate both
computationally and interpretively. Actually, after some
refinements deep tunneling (i.e., well below the top of the
barrier) can also be studied \cite{pollak}. However, for some
other cases, this methodology fails and the problem has to be
solved exactly quantum-mechanically. In these cases, though,
interpretations are often based on classical and semiclassical
arguments. A way to avoid these thought schemes and provide
alternative arguments fully based on quantum-mechanical grounds
consists of considering Bohmian mechanics \cite{bohm,holland-bk}.
The first work in this direction was developed by Hirschfelder
{\it et al.}\ \cite{hirschfelder} in 1974, where tunneling across
a two-dimensional square barrier was studied in terms of
stationary quantum trajectories or quantum streamlines. As it was
shown, under tunneling conditions quantum trajectories present
analogous behaviors to those observed in optics in situations such
as the frustrated total reflection or the Goos-H\"anchen shift.
Later on, in 1982, Dewdney and Hiley \cite{dewdney} analyzed the
problem of time-dependent scattering off square barriers and
wells, formerly considered by Goldberg {\it et al.}\
\cite{goldberg} with wave packets. More recently, Lopreore and
Wyatt started \cite{wyatt1,wyatt2} the development of the
so-called {\it quantum trajectory methods} \cite{wyatt-bk} by
studying tunneling through barriers in one and two dimensions.
Within this methodological schemes, quantum-mechanical problems
are tackled by treating systems as a quantum fluid \cite{madelung}
and then solving the corresponding equations of motion as in
classical fluid dynamics. The wave function or any related
property is then {\it synthesized} from trajectory calculations,
thus skipping a direct use of the time-dependent Schr\"odinger
equation.

An appealing feature to consider Bohmian mechanics in the study of
tunneling dynamics comes from the fact that initial conditions
leading to tunneling can be unambiguously determined
\cite{wyatt1,wyatt2}.
This information can be obtained in a rather practical way.
For example, it allows us to determine how many trajectories pass
the barrier even if at later times they eventually recross it back
and forth again (e.g., problems of chemical reaction dynamics
\cite{sanz-bofill}).
Moved by this fact, here we present an alternative scheme to determine
tunneling conditions and estimate transmission probabilities based on
Bohmian mechanics.
Accordingly, a connection can be established between the transmittance
and the system initial state.
Thus, first we have considered the time-dependent
collision of a Gaussian wave packet with a ramp-like potential, which
can be solved analytically.
Then, from this model, tunneling properties have been inferred for
ramp-like barriers\footnote{In this work, we have only focused on
transmission below the barrier height and therefore important problems
like above-barrier reflection are not considered.
Work related is currently in progress.},
such as those we may find in problems involving electric fields
(e.g., electron ionization processes).
In particular, a connection between the event-to-event fraction of
particles that overcome the barrier and the features characterizing
both the initial wave packet (width and momentum) and the barrier
(slope) has been determined.
Thus, although quantum trajectories are not experimentally
observable\footnote{Recently {\it photon} paths \cite{milena}
have been experimentally inferred \cite{photon-exp} by using weak
measurement processes.}, the information extracted from them could
be used in a practical fashion in the design or characterization of
quantum control experiments involving tunneling (e.g., ionization
processes or chemical reactions).
In this regard, it is worth mentioning that this idea has also been
considered in the literature to determine tunneling times \cite{leavens}
or escape rates for confined multiparticle systems \cite{soffer}.

This work is organized as follows. In section~\ref{sec2}, the
scheme and hypothesis considered to infer tunneling information
from Bohmian trajectories are discussed. In section~\ref{sec3}, a
practical analysis of this information to estimate transmission
probabilities and its check against exact numerical calculations
are presented and discussed. Finally, in section~\ref{sec4}, the
main conclusions extracted from this work as well as its
generalization and potential interest in more complex problems are
summarized.


\section{Tunneling with quantum trajectories}
 \label{sec2}

\begin{figure}
 \begin{center}
 \epsfxsize=7.5cm {\epsfbox{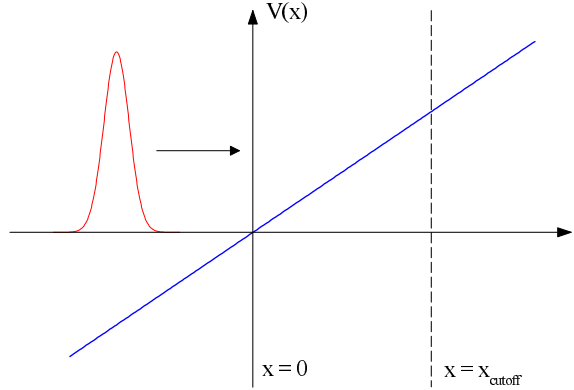}}
 \caption{\label{fig1}
  Schematics of the scattering of a Gaussian wave packet with a linear
  ramp potential.
  If this potential is truncated at some position $x_{\rm cutoff}$
  (i.e., $V(x \ge x_{\rm cutoff}) = 0$), it is shown that tunneling
  may take place depending basically on how the initial wave packet
  is prepared and the uphill slope of the resulting barrier (see
  section~\ref{sec23}).}
 \end{center}
\end{figure}

Square barriers are often considered to carry out time-independent
analysis of tunneling \cite{schiff-bk,liboff-bk}.
However, the collision of a Gaussian wave
packet with a ramp-like barrier can be more easily associated with
realistic tunneling problems, where the ramp represents to some
extent the gradual potential slope met by the wave packet. Thus,
apart from containing the physical elements necessary to
understand quantum processes and phenomena where tunneling is
involved, it is also simple enough to be analytically handled. In
this regard, consider the collision of a Gaussian wave packet with
a linear ramp-like potential, $V(x) = m \alpha x$, as illustrated
in figure~\ref{fig1}. At $t=0$, the wave packet is given by
\be
 \Psi_0(x) = A_0 \ \! e^{- (x-x_0)^2/4\sigma_0^2
  + i p_0 (x - x_0)/\hbar} ,
 \label{eq1}
\ee
where $A_0 = (2\pi\sigma_0^2)^{-1/4}$ is the normalization
constant, $x_0$ and $p_0$ are respectively the (initial) position
and momentum of its centroid (i.e., $\langle \hat{x} \rangle =
x_0$ and $\langle \hat{p} \rangle = p_0 = m v_0$, with $v_0 \ge
0$), and $\sigma_0$ its initial spreading. The ramp has a positive
slope along the $x$-direction (i.e., $\alpha > 0$, since the mass $m$
is always positive).
Thus, after colliding with the potential, the
wave packet will move backwards with an acceleration $a = -
(\partial V/\partial x)/m = - \alpha < 0$.
However, if the ramp is
truncated at some point $x = x_{\rm cutoff}$ (vertical dashed line in
figure~\ref{fig1}), tunneling may appear under certain conditions.
The purpose below is to determine parameters and conditions
leading to this tunneling.


\subsection{Scattering with a linear ramp potential}
 \label{sec21}

As shown elsewhere \cite{sanz-jpa}, the time-evolution of a free
Gaussian wave packet like (\ref{eq1}) is governed by two dynamical
processes: propagation in configuration space and spreading. This
fact becomes apparent through the two terms that appear when
computing the corresponding average energy or energy expectation
value,
\be
 \bar{E} = \langle \hat{H} \rangle = \frac{p_0^2}{2m}
   + \frac{\hbar^2}{8m\sigma_0^2} \equiv \bar{E}_k ,
 \label{eq2}
\ee
where $\hat{H} = \hat{p}^2/2m = -(\hbar^2/2m) \partial^2/\partial
x^2$ and $\bar{E}_k$ stands for the free wave packet average kinetic
energy. Accordingly, a spreading rate or velocity is defined as
$v_s = p_s/m = \hbar/2m\sigma_0$, with $p_s$ being the associated
spreading momentum. The dynamical evolution of (\ref{eq1}) or any
other kind of wave packet regardless of its initial shape (unless
it is spreadless \cite{berry,rau,siviloglou}) is ultimately ruled
by the ratio $v_0/v_s$. This can be easily seen in diffraction
problems, where the ``shape'' of the initial wave function governs
its eventual evolution
\cite{sanz-fractal,sanz-talbot,chia-complex}.

In tunneling processes, the ratio $v_0/v_s$ is also very
important. The time-evolution of the Gaussian wave packet
(\ref{eq1}) colliding with the linear ramp potential of
figure~\ref{fig1} can be straightforwardly obtained analytically
\cite{heller-75,tannor-bk} to yield
\ba
\fl
 \Psi(x,t) & = &
  A_t \ \! e^{- (x-x_{\rm cl})^2/4\tilde{\sigma}_t\sigma_0
   + i p_{\rm cl} (x - x_{\rm cl})/\hbar
   + i (p_0^2/2m - m \alpha x_0) t/\hbar
   - i (p_0 - m \alpha t/3) \alpha t^2/\hbar} \nonumber \\
\fl
 & = & A_t \ \! e^{- (x-x_{\rm cl})^2/4\tilde{\sigma}_t\sigma_0
 + i [ p_{\rm cl} (x - p_0 t/2m) - p_0 x_0 - m \alpha^2 t^3/6]/\hbar} ,
 \label{eq3}
\ea
where $A_t = (2\pi\tilde{\sigma}_t^2)^{-1/4}$ and
$\tilde{\sigma}_t = \sigma_0 (1 + i\hbar t/2m\sigma_0^2)$.
The wave packet (\ref{eq3}) evolves in time without changing its
Gaussian shape, but only increasing its width as
\be
 \sigma_t = |\tilde{\sigma}_t| =
  \sigma_0 \sqrt{1 + \left( \frac{\hbar t}{2m\sigma_0^2} \right)^2 }
  = \sigma_0 \sqrt{1 + \left( \frac{v_s t}{\sigma_0} \right)^2 } .
 \label{eq5}
\ee
Its propagation follows the classical uniform accelerated motion
displayed by its centroid, $x_{\rm cl} = x_0 + v_0 t - \alpha t^2/2$
and $p_{\rm cl} = p_0 - m \alpha t$.
As for its average energy,
\be
 \bar{E} = \bar{E}_k + \bar{V} =
  \frac{p_{\rm cl}^2}{2m} + m \alpha x_{\rm cl}
  + \frac{\hbar^2}{8m\sigma_0^2}
  = \bar{E}_{\rm cl} + \bar{E}_s ,
 \label{eq6}
\ee
where $\bar{V} = m\alpha x_0$.
As it is apparent, this expression
contains the classical-like term $\bar{E}_{\rm cl}$ accounting for
the translation motion and the (quantum-like) spreading energy,
$\bar{E}_s$.
Since there is no coupling between both motions, these two terms remain
constant with time separately.
For more complex potential functions, though, this does not hold; the
``accommodation'' of the wave packet to the corresponding boundary
conditions leads to their eventual nonseparability.
Nevertheless, their initial separability is enough to infer the
long-time dynamics from the earlier stages of the wave packet
evolution \cite{sanz-chemphys}.

Another way to infer the wave packet dynamics is through the
probability density associated with (\ref{eq3}),
\be
 \rho(x,t) = \frac{1}{\sqrt{2\pi\sigma_t^2}} \
   e^{- (x-x_{\rm cl})^2/2\sigma_t^2} .
 \label{eq7}
\ee
The centroid of this function evolves along $x_{\rm cl}$, while its width
spreads at a rate
\be
 \frac{d\sigma_t}{dt} = \frac{v_s^2 t}{\sigma_t} .
 \label{eq8}
\ee
Accordingly, (\ref{eq7}) decelerates at a rate $-\alpha$ as it
approaches the ramp and accelerates later on again at a rate
$\alpha$ after bouncing backwards. The {\it turning point} ({\it
tp}) for its centroid (the position where $v_{\rm cl} = 0$ and the
direction of motion changes) is $x_{\rm tp}^{\rm cl} = x_0 +
v_0^2/2\alpha$, reached at $t_{\rm tp}^{\rm cl} = v_0/\alpha$. From
(\ref{eq8}), we also find that at short times, the spreading
increases uniformly, as $\sim (v_s^2/\sigma_0)t$, which indicates
an {\it acceleration} in the wave packet expansion or ``boost
phase'' \cite{wyatt1}. Hence, a {\it spreading} or {\it boost
acceleration}, $\alpha_s \equiv v_s^2/\sigma_0$, can be defined,
which is different from the {\it dynamical} one, $\alpha$.
In the short-time regime, the wave packet thus spreads as $\sigma_t
\approx \sigma_0 + \alpha_s t^2 /2$, i.e., undergoing a classical-like
uniformly accelerated expansion.
On the contrary, in the long-time regime, the spreading rate is
constant and equal to $v_s$, which means a uniform spreading of
the wave packet, $\sigma_t \approx v_s t$.

The Bohmian trajectories associated with (\ref{eq3}) can be
readily obtained after integrating the (quantum) equation of motion
\cite{bohm,holland-bk}
\be
 \dot{x} = \frac{1}{m} \frac{\partial S}{\partial x} = v_{\rm cl}
  + \frac{\hbar^2 t}{4m^2\sigma_0^2\sigma_t^2} \ \! (x - x_{\rm cl}) ,
 \label{eq9}
\ee
where $S(x,t) = (\hbar/2i) \ln [\Psi(x,t)/\Psi^*(x,t)]$ is the
real-valued phase of $\Psi$.
This yields
\be
 x(t) = x_{\rm cl} + \frac{\sigma_t}{\sigma_0} \ \! \delta_0 ,
 \label{eq11}
\ee
with $\delta_0 \equiv x(0) - x_0$ being the distance between the
initial condition of a quantum trajectory and the centroid initial
position. According to (\ref{eq11}), the relative distance between
any two quantum trajectories $x_1$ and $x_2$ increases with time
as
\be
 \frac{x_2(t) - x_1(t)}{x_2(0) - x_1(0)} =
  \sqrt{1 + \left( \frac{v_s t}{\sigma_0} \right)^2} .
 \label{eq12}
\ee
That is, in agreement with (\ref{eq8}), initially (i.e., short
time-scales) the distance between trajectories increases quadratically
with time ((\ref{eq12}) goes like $\sim \alpha_s t^2/2\sigma_0$),
displaying later on a speed-up provoked by the boost acceleration.
Nonetheless, at even lager times, the trajectories eventually undergo
a slowed-down linear increase with time ($\sim v_s t/\sigma_0$).


\subsection{Calculation of transmission probabilities with quantum
trajectories}
\label{sec22}

In standard treatments of tunneling \cite{schiff-bk,liboff-bk},
transmissions are usually determined from matching conditions that the
wave function and its first derivative have to satisfy at the barrier
edges.
However, in a general, non-analytical
tunneling problem, transmission probabilities are obtained by
computing the quantity
\be
 \mathcal{T}_\infty \equiv \lim_{t \to \infty} \mathcal{T}(t)
  = \lim_{t \to \infty} \int_{x_{\rm cutoff}}^\infty \rho(x,t) dx
 \label{eq19}
\ee
where $\mathcal{T}(t)$ is the {\it restricted probability}
\cite{sanz-sars} in the region behind $x_{\rm cutoff}$.
From a Bohmian viewpoint, (\ref{eq19}) is statistically computed by
considering only those quantum trajectories that reach the region
behind the barrier ($x > x_{\rm cutoff}$) or transmission region
(i.e., as in a standard classical Monte-Carlo sampling).
The initial conditions are randomly distributed according to
$\rho_0(x) = |\Psi_0(x)|^2$ and therefore, at any subsequent time,
the transmitted trajectories will also be randomly distributed
according to the transmitted probability density
$\rho_\mathcal{T}(x,t) \equiv \rho(x > x_{\rm cutoff},t)$.

Consider $x_2^\infty$ denotes the asymptotic position of the first
trajectory that penetrates into the transmission region and
$x_1^\infty$ the position of the last one.
These trajectories can then be identified with the conditions that
define the beginning and the end of the transmitted part of the wave
function at $t \to \infty$.
This allows us to reexpress (\ref{eq19}) as
\be
 \mathcal{T}_\infty
 = \int_{x_1^\infty}^{x_2^\infty} \rho_\mathcal{T} (x^\infty) dx^\infty
 = \lim_{t \to \infty} \int_{x_1(t)}^{x_2(t)}
   \rho_\mathcal{T} (x(t)) dx(t) ,
 \label{eq21}
\ee
where $x^\infty$ denotes the position of a quantum trajectory at
$t \to \infty$ and confined between $x_1^\infty$ and $x_2^\infty$
(i.e., inside the transmission region), $\rho_\mathcal{T} (x^\infty)$
is the probability density evaluated on $x^\infty$ and $dx^\infty$ is
the distance (at $t \to \infty$) between $x^\infty$ and a the
closest neighbor.
The same holds for the expression after the second equality, which
stresses the time-dependence of the quantum trajectories and the
evaluation of the probability density along them (this is why we have
considered explicitly $\rho(x(t))$ instead of $\rho(x,t)$, which simply
expresses the probability density evaluated at a point $x$ at a time
$t$).
This constitutes a very important step.
According to Bohmian mechanics, one can follow the trajectories
backwards in time until reaching their initial conditions, i.e.,
$x_1^\infty \to x_1(0)$ and $x_2^\infty \to x_2(0)$ as $t$ goes from
$\infty$ to 0.
Therefore, (\ref{eq21}) can be recast as an integral over initial
conditions or, in other words, the initial state of the system, as
\be
 \mathcal{T}_\infty = \int_{x_1(0)}^{x_2(0)} \rho(x(0)) dx(0) .
 \label{eq24}
\ee
From this {\it general} result, we find that the knowledge of both the
initial state and the initial conditions $x_1(0)$ and $x_2(0)$, one
can already determine the transmission probability (or get an estimate
of it) performing the whole dynamical calculation.

In order to illustrate the previous assertion, consider the Gaussian
wave packet above.
Substituting (\ref{eq11}) into (\ref{eq7}),
\be
 \rho(x(t)) = \frac{\sigma_0}{\sigma_t} \ \! \rho(x(0)) ,
 \label{eq22}
\ee
and, after assuming small differentials in (\ref{eq7}),
\be
 dx(t) = \frac{\sigma_t}{\sigma_0} \ dx(0) .
 \label{eq23}
\ee
Then, substituting now (\ref{eq22}) and (\ref{eq23}) into (\ref{eq21}),
we obtain (\ref{eq24}).
This is a rather simple way to analytically proof the validity of this
result for the case we are interested in here, namely a Gaussian wave
packet (it will be further worked out later on).
However, let us stress that (\ref{eq24}) it is a general result, which
follows from combining the quantum continuity equation and the
hydrodynamical picture provided by Bohmian mechanics, and has also a
strong connection with the so-called Born rule of quantum mechanics
\cite{brumer} (to some extent, (\ref{eq24}) could be considered as a
consequence of including time-dependence in Born's rule).
Furthermore, despite this physical argumentation, though, a simple
analytical proof can also be found which justifies it\footnote{Numerical
calculations with general wave packets and barriers have also been
carried out to test it, although they are not included in this work.}.
Consider the Jacobian
\be
 \mathcal{J} = \frac{\partial x(0)}{\partial x(t)} ,
 \label{eq23b}
\ee
from which (\ref{eq22}) and (\ref{eq23}) can be expressed in a more
general way as
\be
 \rho(x(t)) = | \mathcal{J} |\ \! \rho(x(0))
  \qquad {\rm and} \qquad
  dx(0) = | \mathcal{J} |\ \! dx(t) ,
 \label{eq23bb}
\ee
respectively.
These expressions stress the fact that, as in classical mechanics, in
Bohmian mechanics there is also a causal connection (mapping) between
two points $x(0)$ and $x(t)$ in configuration space (in classical
mechanics this connection is in phase space).
Therefore, according to a Liouvillian (conservative) viewpoint, a
swarm of initial conditions enclosed within a certain region
$\mathcal{C}$
of the configuration space will end up in another bound region
$\mathcal{C}'$ of this configuration space at a subsequent time,
without any loss of trajectories during the evolution.
Hence, if the integrated probability in $\mathcal{C}$ (i.e., the total
number of trajectories within it) at $t=t_1$ is $P_\mathcal{C}(t_1)$ and,
at some subsequent time $t_2$ the integrated probability in $\mathcal{C}'$
is $P_{\mathcal{C}'}(t_2)$, then $P_{\mathcal{C}'}(t_2) \equiv P_\mathcal{C}(t_1)$ if
the boundary $\mathcal{C}'$ results from the causal evolution of $\mathcal{C}$,
i.e., $\mathcal{C} \to \mathcal{C}'$ when $t_1 \to t_2$.
Similarly, here we have that (\ref{eq19}) and (\ref{eq24}) are
equivalent regardless of the probability density considered, i.e.,
\be
 \int_{x_1^\infty}^{x_2^\infty} \rho_\mathcal{T} (x^\infty) dx^\infty
  = \int_{x_1(0)}^{x_2(0)} \rho_\mathcal{T} (x(0)) dx(0) ,
 \label{eq21proof}
\ee
provided $x_1(0) \to \ x_1^\infty$ and $x_2(0) \to \ x_2^\infty$ as
$t: 0 \to \infty$.

The result (\ref{eq24}) actually holds for any restricted probability
and not only for tunneling transmissions.
For example, in problems such as atom-surface scattering \cite{sanz-prb}
or slit diffraction \cite{sanz-jpcm}, quantum trajectories allow us to
determine unambiguously in the initial probability distribution the
boundaries that separate the contributions that lead to each specific
feature of the corresponding diffraction pattern.
More specifically, consider the so-called {\it peak area intensity},
which is the integral of the probability density covered by a given
diffraction peak.
In the examples mentioned before, the peak area intensity can be
obtained directly from the initial wave function once the two quantum
trajectories that delimit a given diffraction peak are known, just by
computing the number of (quantum trajectory) initial conditions lying
between the initial conditions corresponding to those delimiting
trajectories.
This is exactly the same problem we face here with respect to
tunneling, although instead of dealing with diffraction peak
intensities, we have transmission probabilities.


\subsection{Tunneling through a linear ramp barrier}
\label{sec23}

In order to estimate transmission probabilities from (\ref{eq24}) in
the problem we are dealing with here, first the initial conditions for
the boundary trajectories have to be determined.
In principle, it is reasonable to assume that $x_2(0)$ is such that
$\rho_0 (x_2(0)) \approx 0$, with $\rho_0 (x(0) > x_2(0)) = 0$.
Assuming also that $\rho_0$ must vanish at distances much closer than
$x_{\rm cutoff}$, we can take $x_2(0)$ to infinity for practical
purposes.
On the other hand, $x_1(0)$ will correspond to some $x_0^{\rm min}$ that
constitutes the onset of transmission, i.e., any trajectory such that
$x(0) < x_0^{\rm min}$ will bounce backwards, thus not passing the barrier.
Below, some criteria to estimate $x_0^{\rm min}$ will be given based on
quantum trajectory considerations.
For now, these assumptions allow us to analytically approximate
(\ref{eq24}) by
\be
 \mathcal{T}_\infty \approx \frac{1}{2} \ \!
  {\rm erfc} \left( \frac{x_0^{\rm min} - x_0}{\sqrt{2}\sigma_0}
   \right) ,
 \label{eq26}
\ee
where ${\rm erfc}$ denotes the {\it complementary error function}
\cite{stegun}.
This is a general result provided the initial state is described by a
Gaussian wave packet.

Transmission probabilities can be thus determined from (\ref{eq26}) if
$x_0^{\rm min}$ is known or, alternatively, from (\ref{eq24}) by means of
a random, Monte-Carlo-like sampling of initial conditions, $x_i(0)$,
distributed according to $\rho_0$.
Proceeding in this way, (\ref{eq24}) becomes
\be
 \mathcal{T}_\infty \approx \lim_{\mathcal{N} \to \infty}
  \frac{1}{\mathcal{N}} \sum_{i=1}^\mathcal{N}
   \delta (x_i^{\rho_0}(0) \in (x_0^{\rm min},x_2^\infty)) ,
 \label{eq27}
\ee
where $\delta = 1$ if the initial condition is contained within
the considered spatial range and 0 otherwise ---the sampling
according to the initial probability density is denoted by the
superscript `$\rho_0$'.
In the limit of large $\mathcal{N}$, (\ref{eq27}) approaches
(\ref{eq26}), as already seen elsewhere for chemical reactions and
atom-surface scattering processes \cite{sanz-bofill,sanz-sars}.

Below, we provide some considerations based on the ramp barrier from
figure~\ref{fig1} to determine $x_0^{\rm min}$.
The cutoff is chosen at a distance such that $\rho_0(x_{\rm cutoff})
\approx 0$ in order to ensure there is no probability behind the
barrier; below, we will assume $x_{\rm cutoff} = x_0 + N\sigma_0$,
without loss of generality, where $N$ is related to some probability
onset.
Thus, although (\ref{eq11}) itself is not a solution of the
corresponding tunneling problem, it is very useful in the analysis
of tunneling through such barriers in terms of quantum
trajectories. Depending on the initial translational motion two
cases can be analyzed.


\subsubsection{Tunneling dynamics without initial translational motion.}
 \label{sec231}

When $v_0 = 0$, the wave packet initially rests on the classical turning point and
therefore, at any subsequent time, it will slide down the ramp.
Appearance of tunneling will then be ruled by which term is
dominant in (\ref{eq11}), the dynamical down-hill accelerated motion
induced by the potential or the wave packet spreading. The former
leads the wave packet to move far away from the barrier, while
the latter provokes an anti-downhill motion by `pushing' part
of the wave packet opposite to its translation direction.

For the ramp potential, the centroid classical trajectory
will evolve backwards (toward $x < 0$) since the very beginning and,
in the long-time regime, all quantum trajectories will also move backwards
since their dynamics is ruled by $\alpha$, as seen in (\ref{eq11}).
Moreover, due to the Bohmian non-crossing rule
\cite{sanz-jpa}, apparent from (\ref{eq12}), the distance
among them will increase linearly with time.
Thus, the feasibility of tunneling will rely on the dynamical behavior
undergone by the ensemble of trajectories starting from $x_0$ onward
(with $\delta_0 > 0$) for their path can be eventually
closer to $x_{\rm cutoff}$ and therefore it is likely they tunnel if the
boost is strong enough.
In brief, a quantum trajectory will undergo tunneling if
its {\it turning point} satisfies the condition $x_{\rm tp} \gtrsim
x_{\rm cutoff}$.

The turning point of a
quantum trajectory is given by a vanishing of its
associated velocity (\ref{eq9}), i.e., when the translation
and spreading terms in this relation cancel each other,
making the corresponding (quantum) trajectory to bend over
and evolve backwards.
Equating the left-hand side of (\ref{eq9}) to zero, substituting
(\ref{eq11}) and $v_0 = 0$ into its right-hand side, and then solving
for $t$ the resulting expression eventually yields
\be
 t_{\rm tp} = \frac{\sigma_0}{v_s}\
  \sqrt{ \left( \frac{v_s}{\sigma_0} \right)^4
   \left( \frac{\delta_0}{\alpha} \right)^2 - 1 }
  = \frac{\sigma_0}{v_s}\
  \sqrt{ \left( \frac{\alpha_s}{\alpha} \right)^2
   \left( \frac{\delta_0}{\sigma_0} \right)^2 - 1 } .
 \label{eq13}
\ee
This turning time is given as a function of three
parameters: (1) the slope of the potential or, in other words, the
dynamical acceleration that it imprints on (quantum) particles; (2)
the initial spreading of the wave packet, which is a measure of its
size at the ``preparation'' instant and, therefore, can be somehow
controlled experimentally; and (3) the initial distance between the
corresponding particle and the center of the wave packet, this distance
being inconceivable from the viewpoint of standard quantum mechanics.

Given $\alpha$ and $\sigma_0$, the question is then whether a
particle can cross to the other side of the barrier or not.
This information can be extracted from $\delta_0$ as follows.
It is apparent from (\ref{eq13}) that, in order to observe a ``true''
turning point (i.e., the particle moves forward for a while before
getting backwards), the condition
\be
 \delta_0 > \left( \frac{\alpha}{\alpha_s} \right) \sigma_0
  \equiv \delta_0^c
 \label{eq14}
\ee
must be fulfilled, where we call $\delta_0^c$ the {\it critical
distance} ---equivalently, $x(0) > x_c(0) \equiv
x_0 + \delta_0^c = x_0 + (\alpha/\alpha_s) \sigma_0$.
From this position onwards, all
trajectories will display a turning point different from the
origin (see section~\ref{sec3}), while for any trajectory with
$\delta_0 \le \delta_0^c$ its own initial position $x(0)$ is also
its turning point. The trajectory for which $\delta_0 =
\delta_0^c$ plays the role of a {\it separatrix} or {\it boundary}
between two different dynamical behaviors. Note that $\delta_0^c$
grows very rapidly with $\sigma_0$ ($\sim \sigma_0^4$), which
means that the larger the initial spreading, the further away the
position of the turning point. In other words, turning points will
appear close to $x_0$ in cases of fast-boosting initially-prepared
states, while they will appear far away (even in regions where
$\rho \approx 0$) for slow-boosting states.

The position of the turning point for any trajectory with initial
condition satisfying (\ref{eq14}) is determined by substituting
(\ref{eq13}) into (\ref{eq11}), which yields
\be
 x_{\rm tp} = x_0
  + \frac{\alpha}{2} \left(\frac{\sigma_0}{v_s}\right)^2
  + \frac{1}{2} \left(\frac{v_s}{\sigma_0}\right)^2
   \frac{\delta_0^2}{\alpha}
  = x_0 + \frac{\sigma_0}{2} \frac{\alpha}{\alpha_s}
  + \frac{1}{2} \frac{\alpha_s}{\alpha} \frac{\delta_0^2}{\sigma_0} .
 \label{eq16}
\ee
Note that for $\delta_0 = \delta_0^c$ we
obtain $x_{\rm tp} = x_c(0)$, which is the initial position
for the boundary trajectory.
In order to estimate now which initial positions will lead more likely
to tunneling, we assume they must be close enough to $x_{\rm cutoff}$.
Thus, equating (\ref{eq16}) to $x_{\rm cutoff}$ and the solving the
resulting equation for $\delta_0$, we find
$\delta_0^{\rm min} = (2\alpha\sigma_0/\alpha_s)^{1/2}
[x_{\rm cutoff} - x_0 - (\sigma_0\alpha/2\alpha_s)]^{1/2}$,
%
%
which means that at least all trajectories started from
\be
 x_0^{\rm min} = x_0 + \sqrt{\frac{2\alpha\sigma_0}{\alpha_s}}
  \sqrt{x_{\rm cutoff} - x_0
    - \frac{\sigma_0}{2}\frac{\alpha}{\alpha_s}}
 \label{eq18}
\ee
onward are going to display tunneling through the barrier.
Like $x_c(0)$, $x_0^{\rm min}$ also marks the starting point of
another type of boundary trajectory: a trajectory that sets up a
difference between those with turning points going beyond
$x_{\rm cutoff}$ and those that cannot reach this point.


\subsubsection{Tunneling dynamics with initial translational motion.}
 \label{sec232}

From (\ref{eq11}), when $v_0 \ne 0$ the wave packet moves towards
the ramp and then bounces backwards (its centroid turns at
$t_{\rm tp}^{\rm cl} = v_0/\alpha$, when it is at $x_{\rm tp}^{\rm cl} = x_0 +
v_0^2/2\alpha$). Again, only those trajectories with $\delta_0 >
0$ will have a chance to cross. Nonetheless, now the initial
translational velocity introduces a new control parameter into the
barrier passage process, as seen if we substitute (\ref{eq11})
into (\ref{eq9}), with $v_0 \ne 0$, and then equate to zero the
left-hand side, which renders
\be
 0 = v_0 - \alpha t + \frac{\alpha_s \delta_0}{\sigma_t} \ \! t .
 \label{eq-28}
\ee
Unlike the case discussed in Section~\ref{sec231}, here $v_0 \neq 0$
constitutes a second element that can enhance tunneling.
Equation~(\ref{eq-28}) can be solved for $t$ by using different
techniques employed to solve quartic \cite{nickalls1} and cubic
\cite{nickalls2} equations and different cases can be discussed
(depending on the value of the several parameters involved in it).

First, consider $\sigma_0$ is relatively large and therefore there
is a relatively slow boosting phase. More specifically, the wave
packet spreading is assumed to be negligible approximately up to
$t \sim t_{\rm tp}^{\rm cl}$, which implies a relatively small boost
acceleration compared with $\alpha$. Under this condition, a
first-order expansion in time of (\ref{eq-28}) renders
\be
 0 \approx v_0 - \alpha t + \frac{\alpha_s \delta_0}{\sigma_0} \ \! t ,
 \label{eq-29}
\ee
from which we find
\be
 t_{\rm tp} = \frac{v_0}{\displaystyle \alpha
   - \alpha_s \ \frac{\delta_0}{\sigma_0}} .
 \label{eq-30}
\ee
Of course, this expression is only valid if $\sigma_t \approx \sigma_0$
or $v_0$ is relatively large (in the limit $v_0 \to 0$, it does not
approach the slow-boosting limit of (\ref{eq13})).
Equation (\ref{eq-30}) can be further simplified to
\be
 t_{\rm tp} \approx \frac{v_0}{\alpha}
  \left( 1 + \frac{\alpha_s}{\alpha}\frac{\delta_0}{\sigma_0} \right)
  = t_{\rm tp}^{\rm cl} + t_{\delta_0}
 \label{eq-30b}
\ee
if the second term in the denominator of (\ref{eq-30}) is very small
in comparison with $\alpha$ (e.g., for trajectories very near the
center of the wave packet, for which $\delta_0$ can also be small).
Under this assumption, according to (\ref{eq-30b}), the turning time
increases linearly with $\delta_0$.
Furthermore, since $\sigma_t \approx \sigma_0$, it can be shown that
quantum trajectories will be essentially parallel to the centroidal
one,
\be
 x_{\rm tp} \approx x(0) + (x_{\rm tp}^{\rm cl} - x_0)
  = \delta_0 + x_{\rm tp}^{\rm cl} .
 \label{eq-31bb}
\ee
Therefore, for large $\sigma_0$ (slow boosting), we have
$\delta_0^{\rm min} = x_{\rm cutoff} - x_0 - v_0^2/2\alpha$ and
therefore
\be
 x_0^{\rm min} = x_{\rm cutoff} - v_0^2/2\alpha .
 \label{eq18b}
\ee
That is, regardless of the value of
$\sigma_0$, there is always an onset of tunneling provided the
translational velocity is large enough that the trajectory reaches
the ramp cutoff.

The second case is that of fast boosting, i.e., when the initial
boosting phase occurs much earlier than the trajectory reaches its
turning point.
Assuming $\sigma_t \approx v_s t$ and then substituting this into
(\ref{eq9}) leads to
\be
 0 \approx v_{\rm eff} - \alpha t ,
 \label{eq-32}
\ee
where $v_{\rm eff} = v_0 + v_s \delta_0/\sigma_0$ is an
effective constant velocity encompassing both the translational
and the spreading velocities, and proportional to $\delta_0$.
After (\ref{eq-32}), a quantum trajectory reaches its turning point at
\be
 t_{\rm tp} = \frac{v_{\rm eff}}{\alpha}
  = t_{\rm tp}^{\rm cl} + t'_{\delta_0} ,
 \label{eq-33}
\ee
with $t'_{\delta_0} = (v_s/\alpha \sigma_0) \delta_0$.
As for the turning-point position, we find
\be
 x_{\rm tp} = x_0 + \frac{v_{\rm eff}^2}{2\alpha} ,
 \label{eq-34}
\ee
which looks pretty much like the expression for the classical
turning point, but with $v_0$ substituted by $v_{\rm eff}$. From
(\ref{eq-34}), we obtain
$\delta_0^{\rm min} = (2\alpha\sigma_0/\alpha_s)^{1/2}
(x_{\rm cutoff} - x_0)^{1/2} - v_0 \sigma_0/v_s$ and therefore
%
%
%
\be
 x_0^{\rm min} = x_0 + \sqrt{\frac{2\alpha\sigma_0}{\alpha_s}} \ \!
    \sqrt{x_{\rm cutoff} - x_0} - \frac{v_0}{v_s} \ \! \sigma_0 .
 \label{eq18bb}
\ee
The first part in this relation coincides with (\ref{eq18}) in the
limit $x_{\rm cutoff} - x_0 \gg \sigma_0 \alpha/2 \alpha_s$.
Hence, in the limit $v_0 \to 0$, both cases will approach.
The interest on (\ref{eq18bb}) relies on the fact that, unlike
(\ref{eq18}), if the spreading rate is not enough to overcome
$x_{\rm cutoff}$, there is an extra (translational) energy which may
help to surmount the barrier, this favoring the passage.
The starting point of the quantum trajectory can then be
further away from the barrier cutoff because of the extra velocity.


\section{Numerical results}
 \label{sec3}


\subsection{Parametric study of the transmission probability}
 \label{sec31}

\begin{figure}
 \begin{center}
 \epsfxsize=11cm {\epsfbox{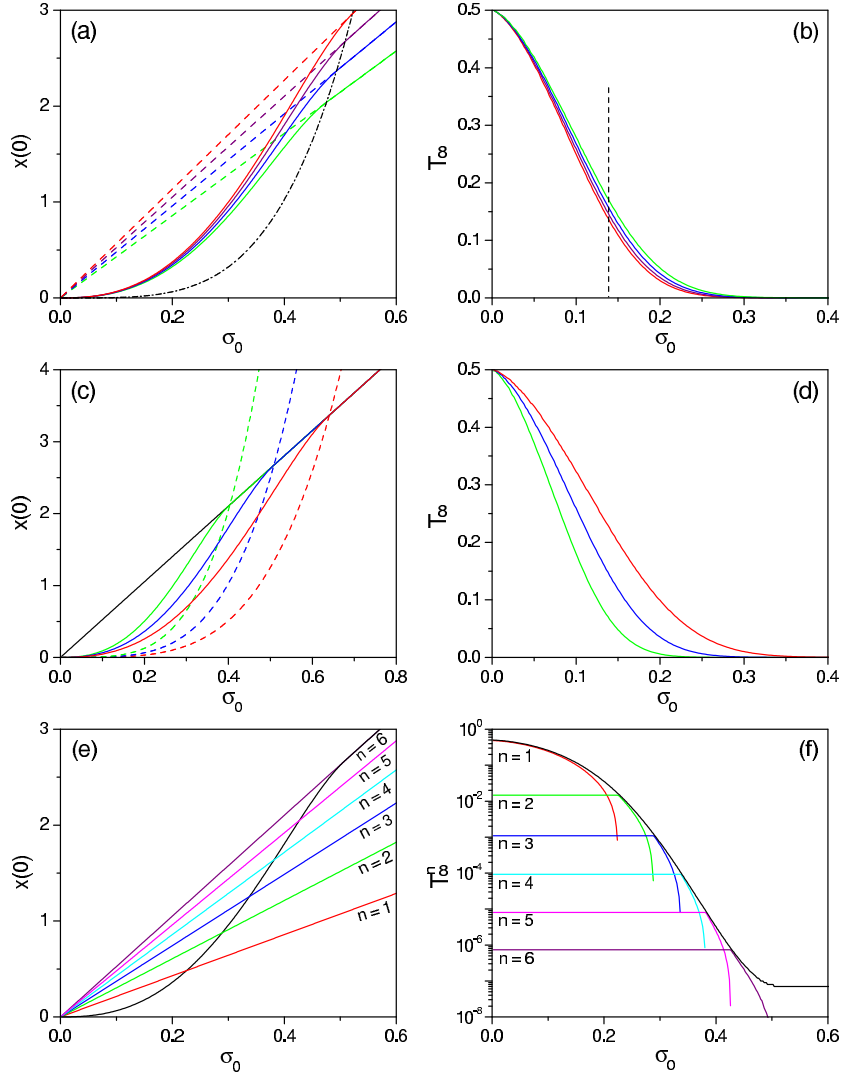}}
 \caption{\label{fig2}
  Top: $\mathcal{T}_\infty$ as a function of $\sigma_0$ for different
  values of the sensitivity parameter $n$ and, therefore, of the cutoff
  position $x_{\rm cutoff}$: $n=4$ (green), $n=5$ (blue), $n=6$
  (purple) and $n=7$ (red).
  In part (a), scheme to choose the initial positions, $x(0)$, for the
  calculation of $\mathcal{T}_\infty$ in panel (b): the cutoff position,
  $x_{\rm cutoff}$, is indicated by the dashed lines, $x_0^{\rm min}$ by
  the solid lines and $\delta_0^c$ by the black dashed-dotted line.
  Center: $\mathcal{T}_\infty$ as a function of $\sigma_0$ for different
  values of the barrier slope: $\alpha = 5$ (green), $\alpha = 10$
  (blue) and $\alpha = 20$ (red).
  In part (c), scheme to choose the initial positions, $x(0)$, for the
  calculation of $\mathcal{T}_\infty$ in panel (d): the cutoff position,
  $x_{\rm cutoff}$, is indicated by the black solid line,
  $x_0^{\rm min}$ by the colored solid lines and $\delta_0^c$
  by the colored dashed-dotted line.
  Bottom: Contributions to $\mathcal{T}_\infty$ from regions with
  different sensitivity range (see text for details).
  In part (e), $x_0^{\rm min}$ is indicated by the thick solid line,
  while the straight lines indicate limit of regions with sensitivities
  up to the value given in the corresponding label.
  In part (f), the thick solid line indicates $\mathcal{T}_\infty$,
  while the colored lines indicate indicate the contributions between
  two consecutive sensitivity values; each color corresponds to the
  upper sensitivity considered from part (e).}
 \end{center}
\end{figure}

In figure~\ref{fig2} we show some calculations\footnote{In all the
calculations presented here, $\hbar = m = 1$. Moreover, when necessary,
quantum trajectories have been calculated using standard wave packet
propagation techniques \cite{sanz-physrep} to obtain the wave function
and then integrating (\ref{eq9}).} of transmission probabilities,
$\mathcal{T}_\infty$, as a function of
different parameters in order to illustrate the concepts
introduced above. Without loss of generality, we have considered
$v_0 = 0$, which allows us to better understand the tunneling
dynamics ``clean'' of possible contributions coming from the
kinetic energy of the particle.

Consider first the effect of varying the cutoff distance,
$x_{\rm cutoff}$. As mentioned above, this distance is chosen as a
function of the initial width, as $x_{\rm cutoff} = x_0 + N \sigma_0$.
To avoid some arbitrariness in the choice of $N$ and ensure
$\rho_0(x)$ vanishes far away from $x_{\rm cutoff}$, we define the
ratio
\be
 \Gamma(x_{\rm cutoff}) \equiv
  \frac{\rho_0(x_{\rm cutoff})}{\rho_0(x_0)} = 10^{-n}
 \label{eq28}
\ee
and call the exponent $n$ the {\it sensitivity parameter}, which
provides us with a minimum bond for the probability density
(relative to its maximum value at the center of the initial wave
packet). For example, setting $n=6$ means that whenever
$\Gamma(x(0)) < \Gamma(x_{\rm cutoff}) = 10^{-6}$, the probability
density will be assumed to be zero (i.e., no probability below
that value will be detected or computed). Accordingly, we have $N
= \sqrt{2n\ln 10} \approx 2.15\ \! n^{1/2}$. In
figure~\ref{fig2}(a) we plot $x_{\rm cutoff}$ (dashed lines) and the
corresponding curve $x_0^{\rm min}$ (solid lines) as a function of
$\sigma_0$ for four different values of the sensitivity parameter:
$n=4$ (green), $n=5$ (blue), $n=6$ (purple) and $n=7$ (red). As it
can be noticed, as $n$ increases, $x_0^{\rm min}$ also increases,
which means to move the integration range in (\ref{eq24}) towards
regions with lower values of $\rho_0(x)$, this leading to smaller
transmissions (see figure~\ref{fig2}(b)). Nonetheless, the most
important changes in the integration range happen at high values
of $\sigma_0$, near the onset of no tunneling, this being the
reason why in figure~\ref{fig2}(b) there is only a difference
$\mathcal{T}_\infty(n=4) - \mathcal{T}_\infty(n=7) \sim 0.0336$
($\sim 6.72\%$ when this difference is referred to 0.5, the
maximum value of $\mathcal{T}_\infty$) between the curves for
$n=4$ and $n=7$, at $\sigma_0 \approx 0.138$ (see the position of
the vertical dashed line). In figure~\ref{fig2}(a) we have also
plotted $\delta_0^c$ (black dashed-dotted line), which does not
depend on the sensitivity parameter.

In the central part of figure~\ref{fig2} the effects of
the slope of the potential on tunneling are displayed for three
different values of the dynamical acceleration: $\alpha = 5$
(green), $\alpha = 10$ (blue) and $\alpha = 20$ (red).
In all cases, the cutoff corresponds to
a sensitivity parameter $n=6$, with the black solid line of
figure~\ref{fig2}(c) denoting the position of $x_{\rm cutoff}$
for each value of $\sigma_0$.
In figure~\ref{fig2}(c), for the same value of $\alpha$, solid
lines indicate the position of $x_0^{\rm min}$ and dashed ones
the position of $\delta_0^c$.
As the barrier slope $\alpha$ increases, tunneling probability
decreases (see figure~\ref{fig2}(d)).
This decrease is again due to shift of the boundary curve $\delta_0^c$
when $\alpha$ increases, which leads to a smaller area between
$x_{\rm cutoff}$ and $x_0^{\rm min}$ as well as to areas where
the value of $\rho_0$ becomes meaningless (see below).
Taking this fact into account, we can distinguish between two limits
of interest.
In the case of $\alpha \to \infty$, i.e., a vertical, infinite wall,
no particle will be able to pass the barrier.
On the contrary, when $\alpha \to 0$ and we have a flat surface, only
half the number of particles from the initial ensemble are going to
pass the barrier.
In this latter case, the maximum value of $\mathcal{T}_\infty$ is only
0.5 because, as can be easily noted, $x_{\rm cl} = 0$ and, therefore, half
of the particles will move forward (those starting in the front part of
the wave packet, for which $\delta_0 > 0$) and another half will move
backward (those from the rear part of the wave packet, with $\delta_0
< 0$).

To understand the different contributions of the region bounded
between $x_{\rm cutoff}$ and $x_0^{\rm min}$ and therefore their role in
the tunneling process, in the bottom part of figure~\ref{fig2}
$\mathcal{T}_\infty$ is plotted as a function of different
sensitivity parameters. In figure~\ref{fig2}(e), we have plotted
$x_0^{\rm min}$ (thick black line) for $\alpha = 10$ and a sensitivity
of $n=6$ (purple line). The positions of $x_{\rm cutoff}$ for
different sensitivities are also plotted. These additional lines
indicate that if $n=2$ (green line), for example, the initial
conditions lying between this line and $x_0^{\rm min}$ will give rise
to probabilities such that $\Gamma(x(0)) \ge 10^{-2}$. The
contribution $\mathcal{T}_\infty^n$ to the transmitted probability
of initial conditions lying between two consecutive sensitivities
$n-1$ and $n$ is represented in figure~\ref{fig2}(f);
$\mathcal{T}_\infty$ is denoted with thick black line and labels
indicate the cutoff chosen, e.g., $n=2$ (green line) represents
$\mathcal{T}_\infty^2$ from trajectories starting between the
cutoffs with sensitivities $n=1$ and $n=2$ (the red and green
lines in figure~\ref{fig2}(e)). As can be seen, the most important
contribution (almost 100\%) arises from trajectories starting
between $x_0^{\rm min}$ and $n=1$, while as $n$ increases the
contributions decrease in about one order of magnitude and are
only relevant as $\sigma_0$ increases.


\subsection{Quantum trajectories and transmission probability}
 \label{sec32}

\begin{figure}
 \begin{center}
 \epsfxsize=12.5cm {\epsfbox{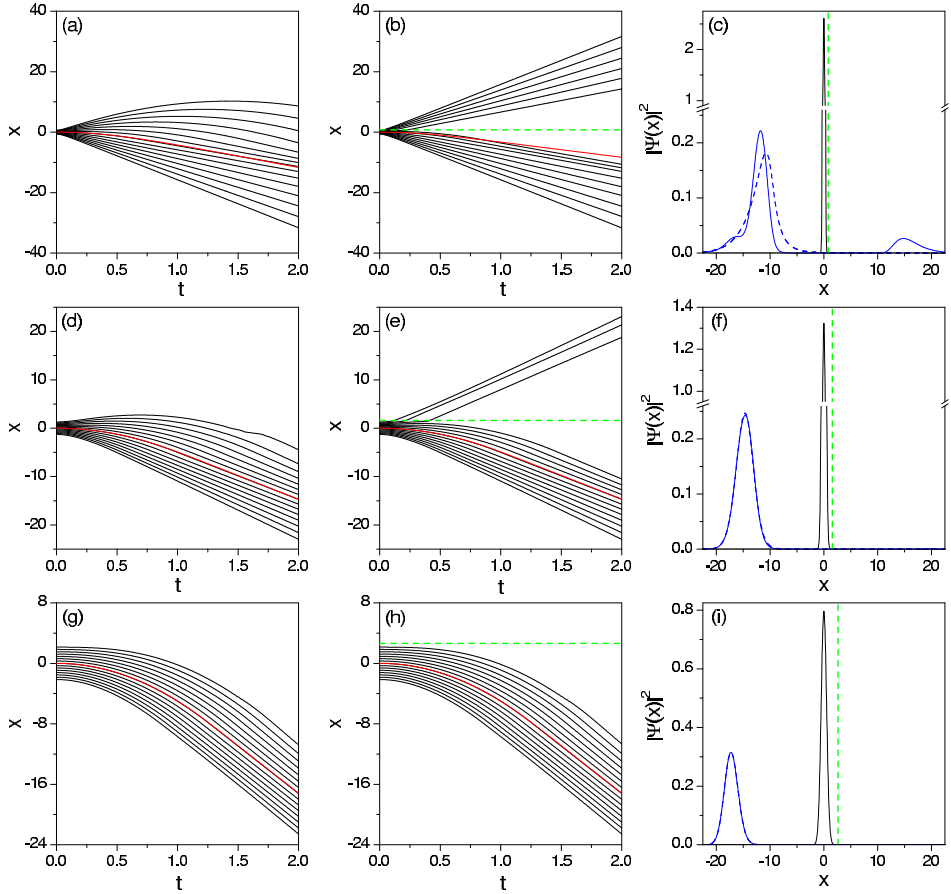}}
 \caption{\label{fig3}
  Quantum trajectories illustrating the collision dynamics with a
  linear, ramp-like potential (left-column panels) and the feasibility
  of tunneling when it is truncated (middle-column panels) for:
  $\sigma_0^{(1)} = 0.15$ (top row), $\sigma_0^{(2)} = 0.3$ (central
  row) and $\sigma_0^{(3)} = 0.5$ (bottom row).
  In each trajectory plot, the thick red curve indicates the average or
  expectation value of the position of the wave packet, $\langle \hat{x}
  \rangle$, with time.
  The position of $x_{\rm cutoff}$ for the corresponding value of
  $\sigma_0$ is shown by the dashed green line (horizontal in the
  middle-column panels and vertical in the right-column ones).
  In the right-column panels, from top to bottom, the initial (black
  line) and final (blue line) probability densities corresponding to
  each value of $\sigma_0$ are displayed.
  In these panels, final probability densities are plotted with solid
  line for the truncated potential and dashed line for the linear,
  ramp-like one.}
 \end{center}
\end{figure}

The results shown above provide us with valuable information
concerning tunneling, although one can go a step further away yet.
In figure~\ref{fig3} different sets of quantum trajectories are
displayed in terms of $\sigma_0$. Thus, from top to bottom:
$\sigma_0^{(1)} = 0.15$ (upper row), $\sigma_0^{(2)} = 0.3$
(middle row) and $\sigma_0^{(3)} = 0.5$ (lower row), which
correspond (see figure~\ref{fig2}(e)) to fast (strong), medium and
slow (weak) boosting phases, respectively. In all cases, $x_0 =
0$, $v_0 = 0$ and $\alpha = 10$. Moreover, for computational
convenience, we have considered a potential function
\be
 V(x) = \left\{ \begin{array}{lcl}
   V_0          & \qquad & x < x_- \\
   m \alpha x   & \qquad & x_- \le x \le x_{\rm cutoff} \\
   V_0          & \qquad & x > x_{\rm cutoff} \\
  \end{array} \right. ,
 \label{eq29}
\ee
where $V_0 = m \alpha (x_0 - 3N\sigma_0)$ and $N = 5.26$, which
corresponds to $n=6$. The left-hand side truncation of the ramp
avoids the increase of downhill velocity of the wave packet; to
minimize effects due to the reflections induced by this change of
slope \cite{schiff-bk,liboff-bk}, we have assumed $x_- = x_0 -
3N\sigma_0$.

In figure~\ref{fig3}(a) we notice that the fast or strong boosting
($\sigma_0 = \sigma_0^{(1)}$) provokes an almost immediate pushing
of some trajectories in the opposite direction to the evolution of
the center of the wave packet (thicker red line). In this case,
the expansion of the wave packet is such that at $t=1$ its width
has increased more than 20 times $\sigma_0$, which is more than
enough for an important number of trajectories to overcome the
barrier (see figure~\ref{fig3}(b)). From a standard quantum
perspective (see figure~\ref{fig3}(c)), the very narrow initial
wave packet (black) passes to a very wide final one (blue dashed
line), which splits into a reflected and a transmitted wave packet
(blue solid lines) when the cutoff is considered (vertical green
dashed line). For a weaker boosting (middle row), the width of the
wave packet at $t=1$ is about 5.6$\sigma_0$ (see
figure~\ref{fig3}(d)). Meanwhile, the cutoff is at $x_{\rm cutoff}
\approx 1.6$, which is only 3.5 times smaller than $\sigma_t$ at
$t=1$. Thus, although some trajectories will be able to cross the
barrier (see figure~\ref{fig3}(e)), they start in regions where
the initial probability density has small values and do not
produce much transmission. As seen in figure~\ref{fig3}(f),
transmission is almost negligible (about 500 times smaller than
reflection). Finally, in the lower row of figure~\ref{fig3}, a
case of slow or weak boosting where quantum trajectories are out
of the conditions leading to tunneling is displayed (see
figure~\ref{fig2}(e))
---the width of the wave packet here is about 2.2 times its
initial width at $t=1$, which is smaller than $x_{\rm cutoff}$.
This is apparent in figure~\ref{fig3}(h), where there are not
transmitted trajectories, and also in Fig.~\ref{fig3}(i), where
there is no transmitted wave packet.

\begin{figure}
 \begin{center}
 \epsfxsize=7cm {\epsfbox{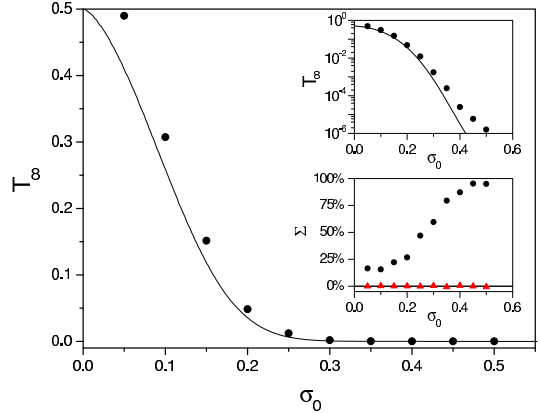}}
 \caption{\label{fig4}
  Transmission probability, $\mathcal{T}_\infty$, as a function of
  the initial width of the wave packet estimated with (\ref{eq26}) and
  (\ref{eq18}) ($\mathcal{T}_\infty^{est}$, solid line) and obtained
  from a wave packet simulation ($\mathcal{T}_\infty^{wp}$, circles)
  for several values of $\sigma_0$.
  In the upper inset, the same plot but at a logarithmic scale to show
  the discrepancy between both calculations as $\sigma_0$ increases.
  In the lower inset, deviation $\Sigma$ (see text for details) for
  the estimated values (black circles) and the corrected ones (red
  triangles) with respect to the values obtained from the wave packet
  simulation.
  The values of the parameters considered are: $\alpha = 10$, $x_0 = 0$
  and $v_0 = 0$.}
 \end{center}
\end{figure}

Finally, in figure~\ref{fig4} we have plotted $\mathcal{T}_\infty$
when it is estimated with the previous simple model
($\mathcal{T}_\infty^{est}$, solid line) and also from wave packet
calculations ($\mathcal{T}_\infty^{wp}$, black circles)
---using quantum trajectories sampled according to $\rho_0$ would
render exactly the same results, as shown elsewhere
\cite{sanz-prb}. As can be noticed, though the estimated
transmission $\mathcal{T}_\infty^{est}$ follows very nicely the
trend of the correct value, $\mathcal{T}_\infty^{wp}$, there is a
discrepancy between both magnitudes which increases with
$\sigma_0$. This can be better appreciated in the upper inset of
the figure, where the $\mathcal{T}_\infty$ axis is given in
logarithmic scale, as well as in the lower inset, where the
deviation between both values (measured in percentages, \%),
\be
 \Sigma \equiv \left( 1 -
  \frac{\mathcal{T}_\infty^{est}}{\mathcal{T}_\infty^{wp}}
  \right) \times 100 \% .
 \label{eq30}
\ee
is calculated. In order to understand these deviations, in
figures~\ref{fig2}(e) and \ref{fig2}(f) we see that, given
$\sigma_0$, those trajectories started between the (red) line for
$n=1$ and $x_0^{\rm min}$ would be the ones with a larger contribution
(the following, between $n=1$ and $n=2$, contribute with a
probability almost two orders of magnitude smaller). Thus, note
that, beyond $\sigma_0 \approx 0.225$ the probabilities are very
low (see green curve in figure~\ref{fig2}(f)) and, therefore,
small variations in the position of $x_{\rm min}$ may imply larger
relative errors, as can be seen in the lower inset of
figure~\ref{fig4}.
Now, the source for these deviations arises from the fact that, in the
model considered here, $x_0^{\rm min}$ was chosen assuming that the
last trajectory penetrating into the transmission region coincides
with $x_{\rm cutoff}$.
This implies to neglect all
those trajectories whose turning points are very close to
$x_{\rm cutoff}$, but without touching it (i.e., with $x(0) \lesssim
x_0^{\rm min}$). Thus, provided they are in a neighborhood of
$x_{\rm cutoff}$, the corresponding trajectories should also be taken
into account and the $x_0^{\rm min}$ curve should be correspondingly
corrected. If we use the trajectory plots to determine the initial
position of the boundary trajectory, and consider it to estimate
again $\mathcal{T}_\infty$ through (\ref{eq26}) but with the new
$x_0^{\rm min}$ positions, effectively, we obtain
$\mathcal{T}_\infty^{wp}$, as indicated by the red triangles in the
lower panel of figure~\ref{fig4}. In order to appreciate the
importance of a good characterization of the region of initial
conditions leading to tunneling, consider, for example, $\sigma_0
= 0.15$. In this case, $x_0^{min,\ est} \approx 0.1774$ and
$\mathcal{T}_\infty^{est} \approx 0.11795$. However, when we
localize the initial position of the boundary trajectory,
$x_0^{min,\ corr} \approx 0.1547$, and correct accordingly the
lower value of initial positions, we obtain
$\mathcal{T}_\infty^{corr} \approx 0.15149$, which is very close
to the value obtained from the simulation,
$\mathcal{T}_\infty^{wp} \approx 0.15149$. Thus, a displacement
$x_0^{min,\ est} - x_0^{min,\ corr} \approx 0.0227$ in an interval
$x_0^{\rm cutoff} - x_0^{min,\ corr} \approx 0.6328$ leads to a
decrease of the true transmission probability of about 22\%, while
the corrected one differs only in about a 0.03\% (both value are
obtained using (\ref{eq30})). Now, if we proceed similarly with
$\sigma_0 = 0.3$, we find $x_0^{min,\ est} - x_0^{min,\ corr}
\approx 0.0823$ in an interval $x_0^{\rm cutoff} - x_0^{min,\ corr}
\approx 0.7024$. This leads to a decrease of the true transmission
probability of about 60\%, while the corrected one differs only
less than 0.5\%.


\section{Conclusions}
 \label{sec4}

Usually in standard quantum mechanics, the transmission or
tunneling probability is associated with the height and width of a
barrier.
Analytical expressions for this probability can be derived
from time-independent or semi-classical calculations
\cite{schiff-bk,liboff-bk} after setting some matching conditions
on both the wave function and its first derivative at the barrier
edges.
On the other hand, from a numerical viewpoint, transmission
probabilities are usually obtained from wave packet calculations,
computing the amount of transmitted probability accumulated beyond
a certain boundary \cite{sanz-bofill}.

A complementary and alternative way to look at tunneling readily
arises when Bohmian mechanics is considered.
In principle, this approach helps us to discern which part of the
incident wave packet is transmitted through the barrier and how this
takes place by monitoring the trajectory flow along time.
However, as we have shown here, one can go beyond these facts and
determine tunneling conditions by studying the individual behavior of
quantum trajectories, which can be used in a practical way to obtain
fair estimates of transmission probabilities with information only
related to the system initial state, according to the general result
(\ref{eq24}).
To illustrate this interesting fact, we have considered the collision
of a Gaussian wave packet with a ramp-like barrier, for which
(\ref{eq24}) can be recast as in the form given by (\ref{eq26}).
This general expression for Gaussian wave packets presents some
remarkable properties.
First, tunneling can be explained essentially in terms of three
physical (measurable) parameters: the wave packet motion (through
$v_0$), the wave packet spreading (through $\sigma_t$ and therefore
$v_s$) and the barrier slope (through $\alpha$).
This arises through the estimates of $x_0^{\rm min}$, according to
expressions like (\ref{eq18}), (\ref{eq18b}) or (\ref{eq18bb}).
Second, once $x_0^{\rm min}$ is set up, (\ref{eq26}) allows us to
estimate and compute transmission probabilities.
Third, conversely, given the transmission probability the inverse
procedure can be used to determine the dynamical boundaries of the wave
packet that lead to tunneling, which can be used to extract valuable
information about the physical properties associated with the wave
packet or the barrier (using expressions like (\ref{eq18}),
(\ref{eq18b}) or (\ref{eq18bb})).

Based on the aforementioned properties, in our opinion, the study
presented here constitutes an important resource at an applied level.
Despite that quantum trajectories are not experimentally observable,
the information they provide can be seen as a tool to analyze
experimental processes and phenomena where tunneling is involved.
In particular, it could be employed to understand and implement
mechanisms aimed at quantum controlling molecular systems, alternative
to (or cooperative with) other mechanisms proposed in the literature
\cite{grossman,batista,arimondo1,arimondo2,ohmura1,ohmura2,lu},
since the treatment here described stresses a direct relationship
between the experimental effect (tunneling) and the initial state.
In virtue of this relationship, fairly well summarized by (\ref{eq24})
or its Gaussian version, (\ref{eq26}), one could control the further
state of the system (i.e., the occurrence of tunneling) by selecting
different values of the parameters involved in the preparation of the
initial state (i.e., the initial conditions of the Gaussian wave
packet).


\ack

Support from the Ministerio de Ciencia e Innovaci\'{o}n (Spain)
under Projects FIS2010-18132 and FIS2010-22082 is acknowledged.
A. S. Sanz would also like to thank the same Institution for a
``Ram\'on y Cajal'' Research Fellowship.


\end{document}